\documentclass[conference]{IEEEtran}
\usepackage{ifpdf}

\usepackage{cite}

\ifCLASSINFOpdf
   \usepackage[pdftex]{graphicx}
\else
\fi
 \usepackage{graphicx}
\usepackage{epstopdf}
\usepackage{lipsum}

\usepackage[cmex10]{amsmath}
\usepackage{algorithmic}

\usepackage{array}

\usepackage{mdwmath}
\usepackage{mdwtab}

\usepackage{eqparbox}

\usepackage[tight,footnotesize]{subfigure}

\usepackage[caption=false]{subfig}
\usepackage{multirow}

\usepackage{fixltx2e}

\usepackage{stfloats}

\usepackage{url}

\begin{document}
%
\title{Multi-RATs Support to Improve V2X Communication}

\author{\IEEEauthorblockN{Ji Lianghai, Andreas Weinand, Bin Han, Hans D. Schotten}
\IEEEauthorblockA{Chair of Wireless Communication, University of Kaiserslautern, Germany, $\lbrace$ji,weinand,binhan,schotten$\rbrace$@eit.uni-kl.de}}


%



\IEEEoverridecommandlockouts
\IEEEpubid{\makebox[\columnwidth]{\copyright~Copyright 2017
		IEEE \hfill} \hspace{\columnsep}\makebox[\columnwidth]{ }}
\maketitle

\begin{abstract}
As the next generation of wireless system targets at providing a wider range of services with divergent QoS requirements, new applications will be enabled by the fifth generation (5G) network. Among the emerging applications, vehicle-to-everything (V2X) communication is an important use case targeted by 5G to enable an improved traffic safety and traffic efficiency. Since the V2X communication requires a low end-to-end (E2E) latency and an ultra-high reliability, the legacy cellular networks can not meet the service requirement. In this work, we inspect on the system performance of applying the LTE-Uu and PC5 interfaces to enable the V2X communication. With the LTE-Uu interface, one V2X data packet is transmitted through the cellular network infrastructure, while the PC5 interface facilitates the direct V2X communication without involving the network infrastructure in user-plane. In addition, due to the high reliability requirement, the application of a single V2X transmission technology can not meet the targets in some scenarios. Therefore, we also propose a multi-radio access technologies (multi-RATs) scheme where the data packet travels through both the LTE-Uu and PC5 interfaces to obtain a diversity gain. Last but not least, in order to derive the system performance, a system level simulator is implemented in this work. The numerical results provide us insights on how the different technologies will perform in different scenarios and also validate the proposed multi-RATs scheme.
\end{abstract}


%
\IEEEpeerreviewmaketitle

\section{Introduction}\label{intro}
%
%
In Europe alone, around 40 000 people die and 1.7 million are injured annually in traffic accidents. At the same time, traffic increases on our roads leading to traffic jams, increased travel time, fuel consumption and increased pollution \cite{D11}. Cooperative intelligent traffic systems (C-ITS) can address these problems by warning drivers of dangerous situations and intervene through automatic braking or steering if the driver is unable to avoid an accident. Besides, cooperative driving applications, such as platooning (road-trains) and highly automated driving can reduce travel time, fuel consumption, and $\text{CO}_2$ emissions and also increase road safety and traffic efficiency. The C-ITS systems rely on timely and reliable exchange of information among traffic participants.\\
In order to provide a wireless network to enable the information exchange process, the next generation of the cellular network (5G) should be designed to offer a solution with a high degree of reliability and availability\cite{D11}\cite{D211}. The information exchange process is often referred to as vehicle-to-everything (V2X) communication. What is expected for V2X communication in 5G is 5 times reduced end-to-end (E2E) latency with much higher reliability compared with the current 4G network \cite{D62}.\\
To provide solutions to V2X communication, a lot of work in the literature focus on the exploitation of direct V2X communication \cite{3gpp}\cite{survey}\cite{medhoc}, where the data packets are directly transmitted from the transmitter to the receiver without going through the network infrastructure. For instance, the 3rd Generation Partnership Project (3GPP) proposes to use PC5 interface to facilitate the direct communication between the two ends of V2X communication \cite{23303}. With the proposed direct V2X communication, the transmission latency can be efficiently reduced, since network infrastructure is not involved in the data transmission. Apart from that, the communication taking place directly between two user devices can also contribute to a higher system capacity for mobile broadband services \cite{vtc2017} or a better power usage for massive machine type communications \cite{commag}. Meanwhile, it is worth noticing that, the evolution of the legacy 4G network (i.e., LTE network) is also targeting at providing V2X communication by transmitting data packets through the cellular network infrastructure \cite{23785}. This transmission technology relies on the LTE-Uu radio interface with some technical enhancements (e.g., a short frame length).\\
As mentioned before, the ultra-high reliability requirement of V2X communication poses a big technical challenge, therefore exploiting one single radio access technology (RAT) (i.e., either LTE-Uu or PC5) is not always sufficient to support V2X services. For example, in order to enhance the perception of the environment to avoid accidents, it is proposed in \cite{22886} that a reliability of 99.99\% should be supported for message transfer among a group of V2X users (UEs) within a communication range of 500 meters. Thus, a multi-RATs diversity scheme is proposed in this work to improve system performance.\\
\begin{figure}[b]
\centering
\includegraphics[scale=0.30]{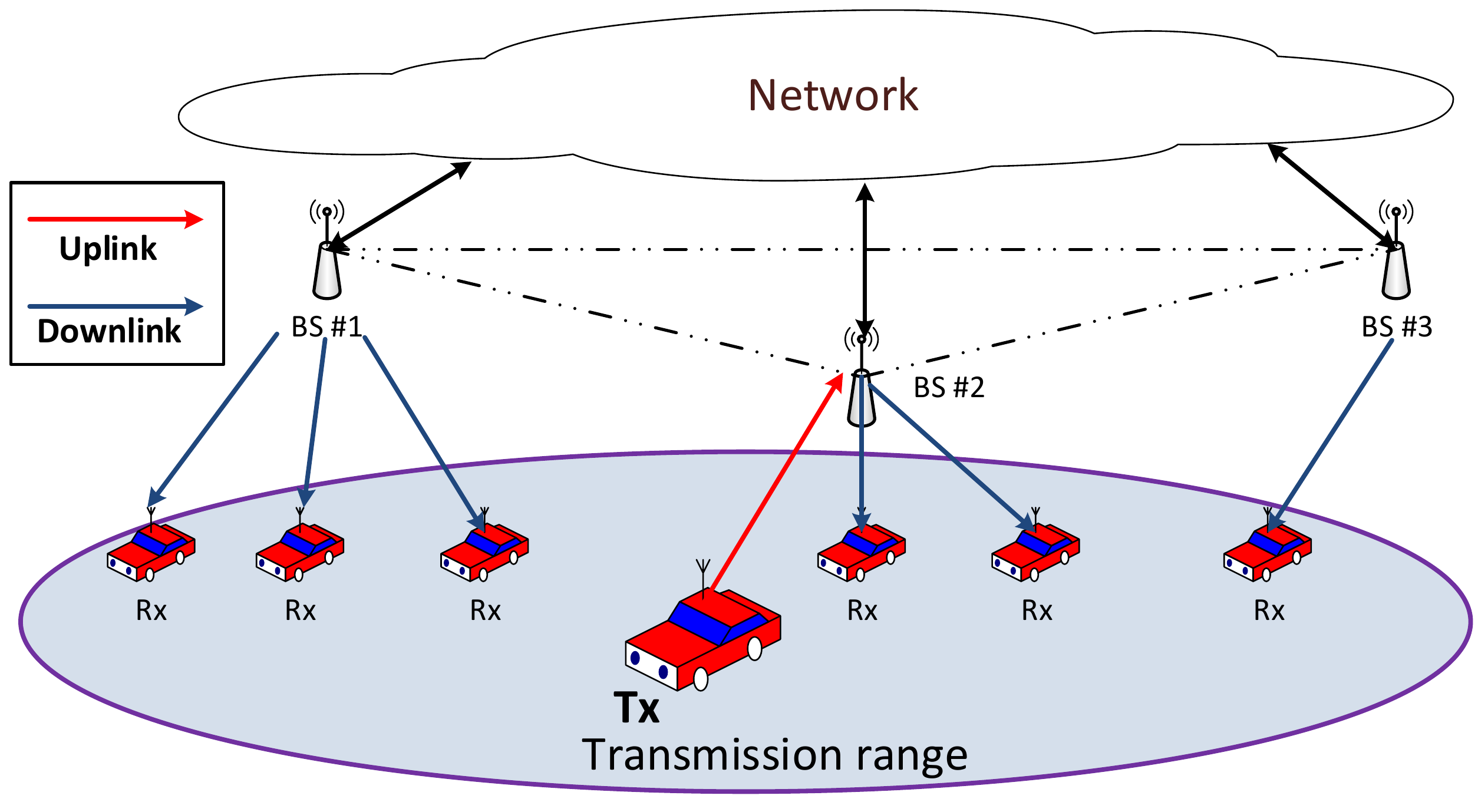}
\centering\caption{V2X communication through network infrastructure}
\label{Uu}
\end{figure}
In this work, we at first inspect on applying the LTE-Uu and PC5 interfaces to enable V2X communication. In Sect.~\ref{LTE-C}, the details on applying the LTE-Uu interface for V2X communication are provided where data packets travel through the cellular network infrastructure. In Sect.~\ref{sidelink}, the V2X communication takes place directly between two nearby UEs over the PC5 interface which is specifically designed for the proximity service in 3GPP. In order to improve the reliability of the V2X communication, a multi-RATs scheme with data packets transmitting through both the LTE-Uu and PC5 interfaces is described in Sect.~\ref{diversity}. Following that, the system models used to evaluate the different technologies are described in Sect.~\ref{SR} and the simulation results will also be provided to demonstrate the performances of the different transmission technologies. Finally, the conclusion of our work is drawn in Sect.~\ref{con}.
\section{V2X Communication through Network Infrastructure}\label{LTE-C}
Fig.~\ref{Uu} demonstrates the case where V2X communication is performed over the LTE-Uu interface. LTE-Uu interface is the air interface between eNodeB and its UEs in LTE network. With this interface, the transmitter (Tx) vehicle sends its data packets to its serving base station (BS) in uplink. And upon the successful reception at the BS, the network further forwards the packets to the relevant receivers (Rxs) in downlink. In order to assist the automatic driving, each V2X data packet is intended to be received by the users in the proximity of the Tx. Therefore, if these users are served by different BSs, the transmission of a packet in downlink involves more than one BS. As shown in Fig.~\ref{Uu}, the network needs to route the received message from BS \#2 to the BSs \#1 and \#3, in order to distribute the received packets to all relevant Rxs. Based on the above statement, the user-plane (UP) end-to-end (E2E) latency, which is the one way transmission time of a packet between the Tx and the Rx, composes of three components, as follows.
\begin{itemize}
\item Uplink latency - the time difference between the generation of a packet at the Tx and its successful reception by the serving BS.
\item Propagation latency between BSs - the packet propagation time between the serving BS of the Tx and the serving BS of a relevant Rx.
\item Downlink latency - the time difference between the successful packet arrival at the serving BS of the Rx and its successful arrival at the Rx.
\end{itemize}
In the rest of this section, we provide more insights into these components. Please note that the E2E latency in this work is inspected in the UP and therefore the control-plane (CP) functions (e.g., radio resource control (RRC) state transition, system information acquisition, and mobility control) are not considered here.
\subsection{Uplink transmission latency}
The uplink transmission refers to a point-to-point (P2P) communication process. The algorithm to evaluate the uplink latency by applying the LTE-Uu interface is presented in \cite{C-V2X}. 
\subsubsection{Scheduling schemes}\label{SS}
Before a UE starts its transmission in the uplink, it needs to obtain the scheduling information regarding the time-and-frequency resource for the transmission. In LTE network, two scheduling schemes can be applied for the network to assign transmission resource for V2X communication. In the first case, the Tx needs to dynamically send a scheduling request (SR) message to the BS, once a data packet arrives at the buffer of the Tx. After that, the network will schedule certain resource and send this configuration information in the downlink control information (DCI) back to the Tx. With this information, the Tx can find the time-and-frequency resource location for its transmission. This scheduling scheme is not effective for certain V2X communication scenarios, taking into account of its specific traffic pattern. For instance, a typical scenario in V2X communication is that each vehicle periodically transmits its data packets to its surrounding vehicles. And the packet generation frequency can be in a range of 1Hz~-~40Hz depending on the applied scenario. In this case, a high signaling overload will be introduced with the dynamic resource request procedure, due to the frequent request of transmission resource. Alternatively, the other scheduling scheme applies a semi-persistent scheduling (SPS). In this scheme, the network assigns a set of time-periodic resources to the V2X Tx. Therefore, once the Tx obtains the SPS configuration information from the BS, it can periodically transmit its data, without triggering a new SR procedure. Compared with the dynamic scheduling approach, the SPS approach has a better support for the periodical traffic of V2X communication due to its less signaling overhead.
\subsubsection{HARQ transmission}
As pointed out before, a P2P communication is applied in uplink. Thus, either an acknowledgment (ACK) or a non-acknowledgment (NACK) message can be sent back from the Rx to the Tx, depending on whether the packet transmission is successful or not. In case a packet reception is failed at the Rx, the NACK message triggers a hybrid-ARQ (HARQ) retransmission procedure. And the Rx can utilize both the previously received packet and the retransmitted packet to correct the error. It is worth noticing that, according to the protocol of LTE, the minimal time interval between the end of a packet transmission and the start of its retransmission is set to be 7 ms, if the retransmission is triggered by a NACK message. In order to meet the low latency requirement, it is currently under consideration of 3GPP to reduce this minimal retransmission interval.
\subsubsection{Modulation and coding scheme}
In order to guarantee a robust transmission, a modulation and coding scheme (MCS) with a low spectral efficiency is required. In LTE system, if the channel state information (CSI) is available at the Tx, it selects the MCS which has the highest spectral efficiency and meanwhile can offer a block error rate (BLER) lower than 10\%. The consideration behind LTE is to find a good compromise between spectral efficiency and robustness, w.r.t. the human-driven traffic. Due to the ultra-high reliability requirement in the emerging V2X communication, the selection of an MCS for V2X communication should be optimized w.r.t. the V2X service requirements. 
\subsection{Propagation latency between the relevant BSs}
As shown in Fig.~\ref{Uu}, the uplink received packets need to be routed through the network to the relevant BSs before the downlink transmission. This procedure refers to a point-to-multipoint (P2MP) transmission in the core network of an operator. An efficient way to realize this P2MP transmission is to apply the evolved multimedia broadcast multicast services (eMBMS). eMBMS is able to provide efficient delivery of broadcast and multicast services in the core network. Compared with the system architecture used for the P2P transmission, two additional components (i.e., Broadcast Multicast - Service Centre (BM-SC) and MBMS-GW in \cite{23468}) are deployed in the core network. The BM-SC is responsible for the management of the eMBMS service-related information, e.g., mapping the service information to the QoS parameters. And the MBMS-GW is the element to deliver eMBMS traffic to multiple cell sites. Considering the Rxs of a V2X packet are the traffic participants in the geometrical proximity of the Tx, it is proposed in 3GPP to localize certain functional entities of the eMBMS architecture to be at the edge of the radio access network (RAN) \cite{23785}.
\subsection{Downlink transmission latency}\label{DL}
In downlink, there are two options to realize the V2X communication. One option is to apply unicast transmission, where a packet is in parallel transmitted to multiple Rxs. For instance, as shown in Fig.~\ref{Uu}, the BS \#1 needs to send one data packet for three times, in order to deliver this packet to the three different Rxs. However, this approach is not efficient for V2X communication in many cases. For instance, during the busy hours of a day, a high density of traffic participants is expected, and therefore the unicast transmission in downlink can easily overload the network. To solve this problem, exploitation of a multicast transmission (i.e., eMBMS in LTE) provides an efficient alternative to V2X communication in downlink.\\
As a specification of the LTE-Uu interface, eMBMS is applied to deliver the same content to multiple Rxs. Therefore, the P2MP scheme offers a high efficiency for transmission of common content. However, in order to successfully deliver a packet to multiple Rxs, eMBMS needs to apply an MCS taking into account of the Rx with the worst radio channel condition. In this sense, a compromise between efficiency and robustness needs to be achieved for the eMBMS. Another character of the eMBMS is the absence of the feedback channel. This means that there is no ACK/NACK message sent back from the Rx to the Tx. Otherwise, the feedback messages from a group of Rxs will introduce a large signaling overhead. In order to enhance the reliability of the multicast transmission in downlink, the BS can try to repeat the transmission of the same data packet, if there is enough resource available in the system. In this way, a maximal ratio combining (MRC) process can be carried out at the Rx. For instance, if the receiver \#n is trying to receive the packet \#m in its $k$-th trial, it will combine all the received copies of this packet. Thus, the signal-to-interference-plus-noise ratio (SINR) can be calculated as:
\begin{equation}
SINR^{\text{MRC}}(m,n,k)=\sum_{i=1}^{k}SINR(m,n,i).
\end{equation}
The term $SINR^{\text{MRC}}(m,n,k)$ represents the post-MRC-processed SINR value of the transmitted packet \#m at receiver \#n after receiving the $k$-th packet copy. And $SINR(m,n,i)$ is the pre-MRC-processed SINR value of the packet \#m at receiver \#n for the $i$-th received copy. As can be seen from this equation, the SINR values can be efficiently increased by utilizing the multiple received copies of a packet.
\begin{figure}[!t]
\centering
\includegraphics[width=3.3in]{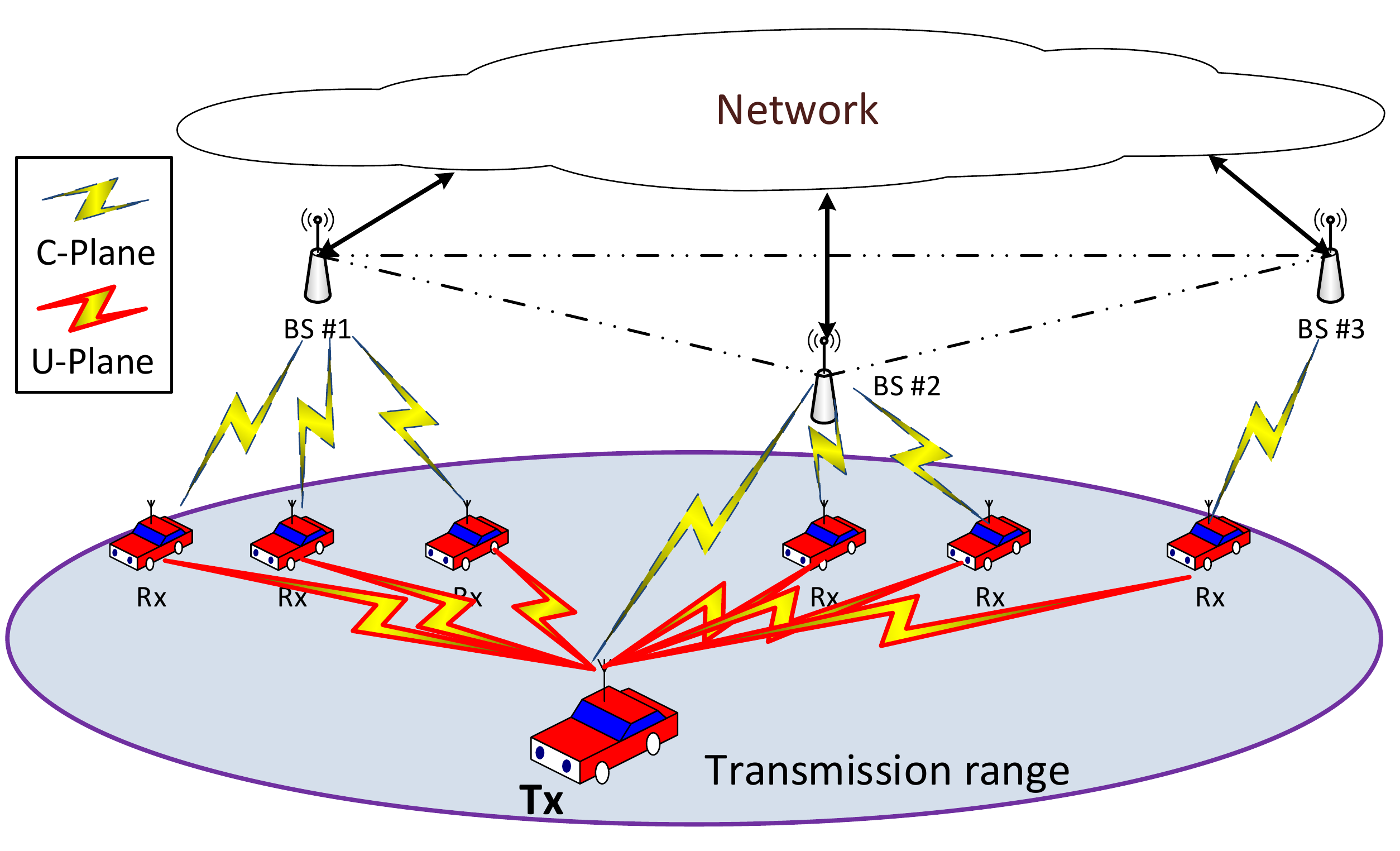}
\caption{Direct V2X communication with network assistance}
\centering
\label{direct}
\end{figure}
\section{Direct V2X Communication}\label{sidelink}
As mentioned in Sect.~\ref{intro}, the V2X communication refers to a local information exchange procedure and therefore the relevant Rxs are the ones located in the proximity of the Tx. Therefore, a packet transmission through network infrastructure is not efficient from the latency perspective. With this motivation, a direct V2X communication can be applied. In this scheme, a Tx directly transmits its data packets to the surrounding Rxs, as shown in Fig.~\ref{direct}. Please note that the UEs can establish a CP connection to the network over LTE-Uu and the direct V2X communication is only in UP.\\
In 3GPP, the direct link between two UEs is referred to as sidelink and the applied air interface is the PC5 interface \cite{23303}. Up to 3GPP release 14, the sidelink communication is connection-less which means there is no RRC connection over PC5. Thus, a V2X Tx multicasts its data packets to the surrounding Rxs and a Rx will discard the received packet if it locally finds the packet is irrelevant. Currently, there are two modes defined in 3GPP to assign radio resource to V2X sidelink transmission. The sidelink transmission mode 3 corresponds to a sidelink transmission over the resource scheduled by the network. In this case, a V2X Tx needs to be in the RRC\_connected state and it sends a Sidelink UE Information message to the BS for resource request. The network then schedules certain resource and sends this information to the V2X Tx in the DCI. It is worth mentioning that the SPS introduced in Sect.~\ref{SS} can also be applied for the sidelink communication. To assist the resource allocation in the BS, UE context information (e.g., traffic pattern, geometrical information) can be collected. For V2X UEs in RRC\_idle state and V2X UEs out of cellular network coverage, the sidelink transmission mode 4 can be applied. In this mode, the information regarding the resource pools of sidelink communication is either broadcasted in the system information blocks (SIBs) \cite{36300} or pre-configured in the V2X UEs. After the acquisition of this information, a V2X Tx in mode 4 autonomously selects a resource from the corresponding resource pool.\\
As mentioned before, the V2X transmission over PC5 is in a multicast manner and therefore there is no HARQ feedback from the Rxs to the Tx. In order to improve reliability, the identical copies of the original transmission can be sent and a Rx performs soft-combining of all received copies to conduct decoding.
\section{Reliability Improvement by Applying Transmission Diversity}\label{diversity}
The design of V2X communication should meet the requirements in the real world. In \cite{22886}, the requirements of different V2X use cases are given. In this document, requirements of different key performance indicators (KPIs) (e.g., the target E2E latency, reliability, and communication range) are proposed. For instance, in order to facilitate a fully automated driving, V2X communication with a reliability value of 99.99\% within a range of up to 500 meters is required. The reliability is defined here as the packet reception ratio (PRR) within the latency requirement. Later in Sect.~\ref{SR}, where system performance will be provided, we will see it is challenging to fulfill the high reliability requirement by applying a single RAT. In this section, we inspect on how to improve the system performance by applying a multi-RATs scheme.\\
\begin{figure}[t]
\centering
\includegraphics[width=3.4in]{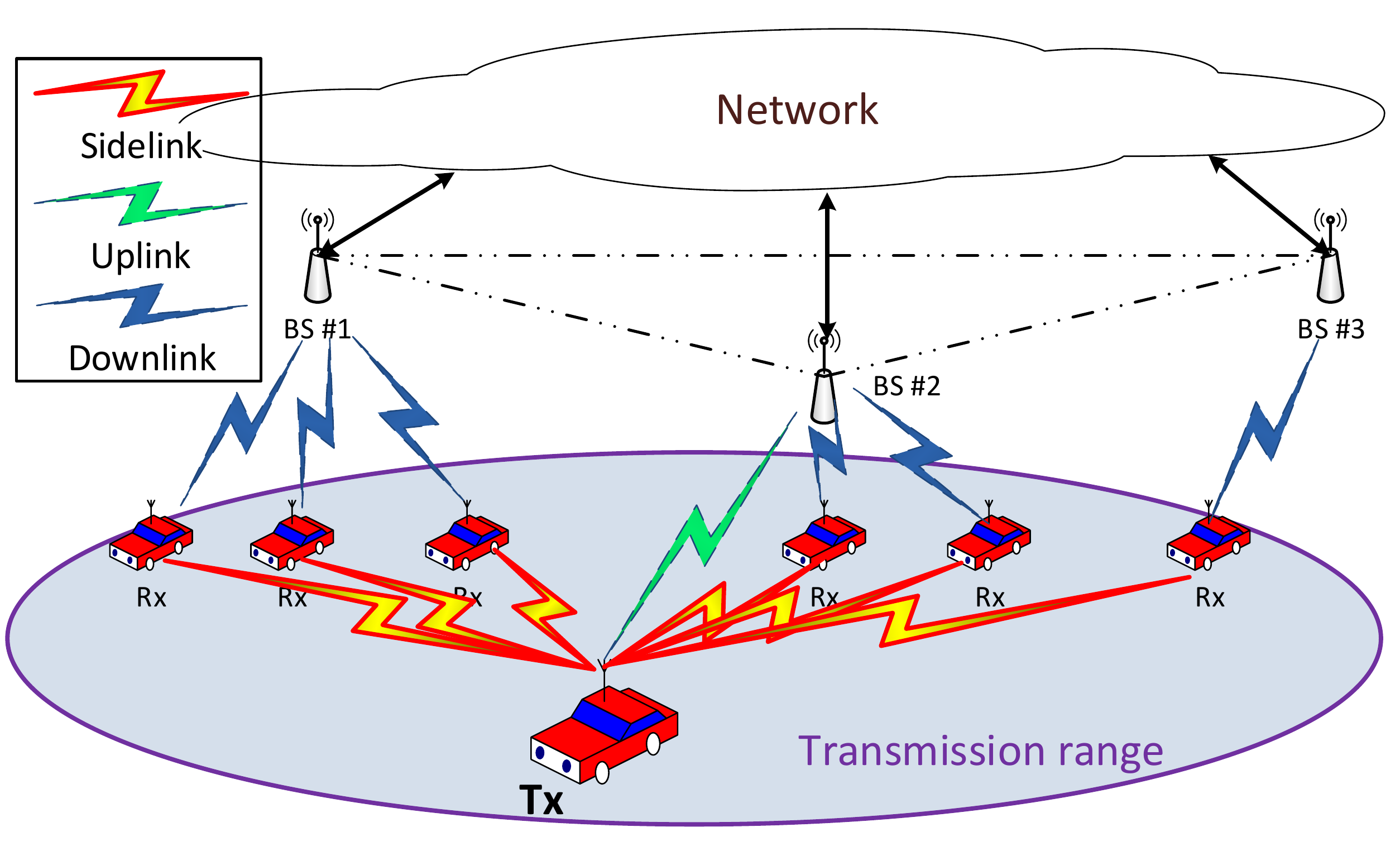}
\caption{Transmission of the V2X packet through both LTE-Uu and PC5}
\centering
\label{diversity_figure}
\end{figure}
Fig.~\ref{diversity_figure} demonstrates the proposed reliability enhancement scheme where a V2X message with the ultra-reliable low-latency requirement is transmitted through both the LTE-Uu and PC5 interfaces. As mentioned in Sect.~\ref{sidelink}, the V2X Tx can transmit the packet over PC5 in both RRC\_connected (i.e., sidelink transmission mode 3) and RRC\_idle states (i.e., sidelink transmission mode 4). However, in order to send the SR message to the BS for uplink resource acquisition, the V2X Tx needs to be in the RRC\_connected state. Thus, once a new V2X packet arrives at the buffer of the Tx, the V2X transmission both over PC5 and LTE-Uu can be performed as:
\begin{itemize}
\item For uplink transmission over LTE-Uu, the Tx needs to send an SR message to the BS and then the BS reserves and configures certain resource for the uplink transmission. After the BS successfully receives a packet, it will further forward this packet to other relevant BSs and then sent it in downlink to the respective Rxs.
\item If sidelink transmission mode 3 is used, the Tx sends a Sidelink UE information message to the BS to request sidelink resource. Once the Tx gets the scheduling information from the BS, it multicasts the data packet over the PC5 interface.
\item If sidelink transmission mode 4 is used, the V2X Tx autonomously selects the resource and transmits the packet over the selected resource.
\end{itemize}
Please note that the above procedure takes place when the V2X Tx needs to obtain the resource configuration before its transmission. In other words, it corresponds to the cases where dynamic scheduling approach is applied or the resource configured by the SPS is not valid anymore. In addition, as mentioned in Sect.~\ref{SS}, the SPS can be applied to reduce the signaling overhead. With this approach, the network configures the transmission resources for once and then the Tx can transmit its packets periodically. Specifically, considering the proposed multi-RATs scheme, the SPS needs to jointly configure the transmission resources for both the LTE-Uu and PC5 interfaces. In this work, the sidelink transmission mode 3 is applied. 
\section{Evaluation Methodology and Numerical Results}\label{SR}
\subsection{System models}
In this work, a dense urban scenario \cite{D61} is used here for simulation purpose and the detailed simulation assumptions of this scenario can be found in \cite{D61}.
\begin{figure}[t]
\centering
\includegraphics[width=3.3in]{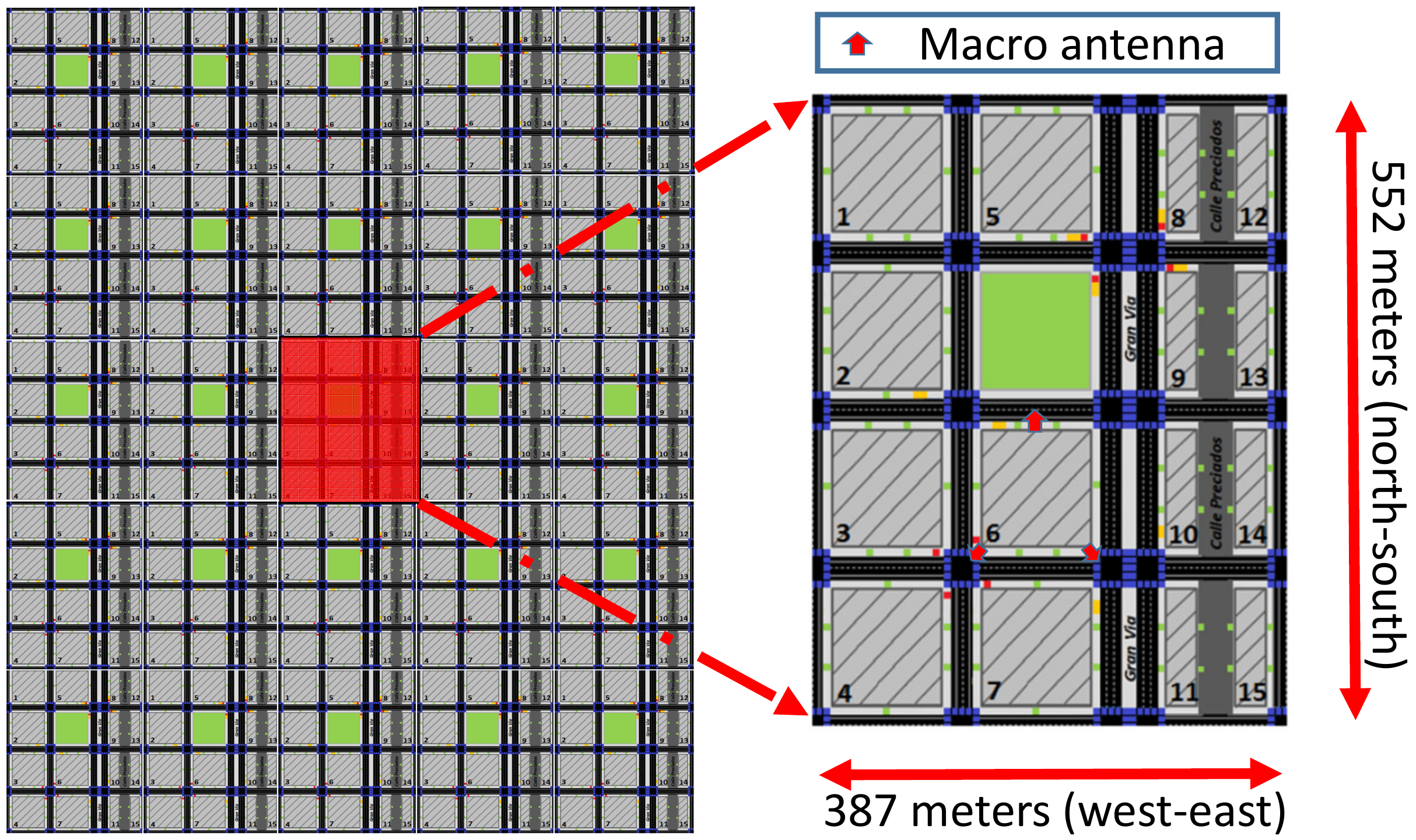}
\caption{Environment model}
\centering
\label{model}
\end{figure}
\subsubsection{Environment model}
In order to characterize the real world environment, a Madrid-grid shown in Fig.~\ref{model} is implemented as the environment model. In this model, each Madrid-grid (i.e., colored as red) is composed of 15 buildings and one park.
\subsubsection{Deployment model}
The system operates on a carrier frequency of 2 GHz and 20 MHz bandwidth (10 MHz/10 MHz for uplink/downlink) is used for the LTE-Uu interface. Additionally, a 10 MHz bandwidth on the carrier frequency of 5.9 GHz is dedicated for the PC5 interface. A macro base station with three sectors is deployed on the roof of the central building, as shown in Fig.~\ref{model}. Besides, an isotropic antenna is installed on each vehicle at 1.5-meter height and each vehicle has a constant transmission power of 24 dBm in each 10 MHz bandwidth.
\subsubsection{Traffic model}
A packet of 212 bytes is generated with 10 Hz periodicity \cite{packet}.
\subsubsection{Channel model}
Line-of-sight (LOS) propagation \cite{D61} and non-line-of-sight (NLOS) propagation \cite{NLOS} are both modeled for V2X communication.
\subsubsection{Mobility model}
A maximal velocity of 50 km/h is assumed, which corresponds to the maximal velocity allowance in most of the European cities.
\subsection{Simulation results}
\begin{figure}[b]
\includegraphics[width=3.4in]{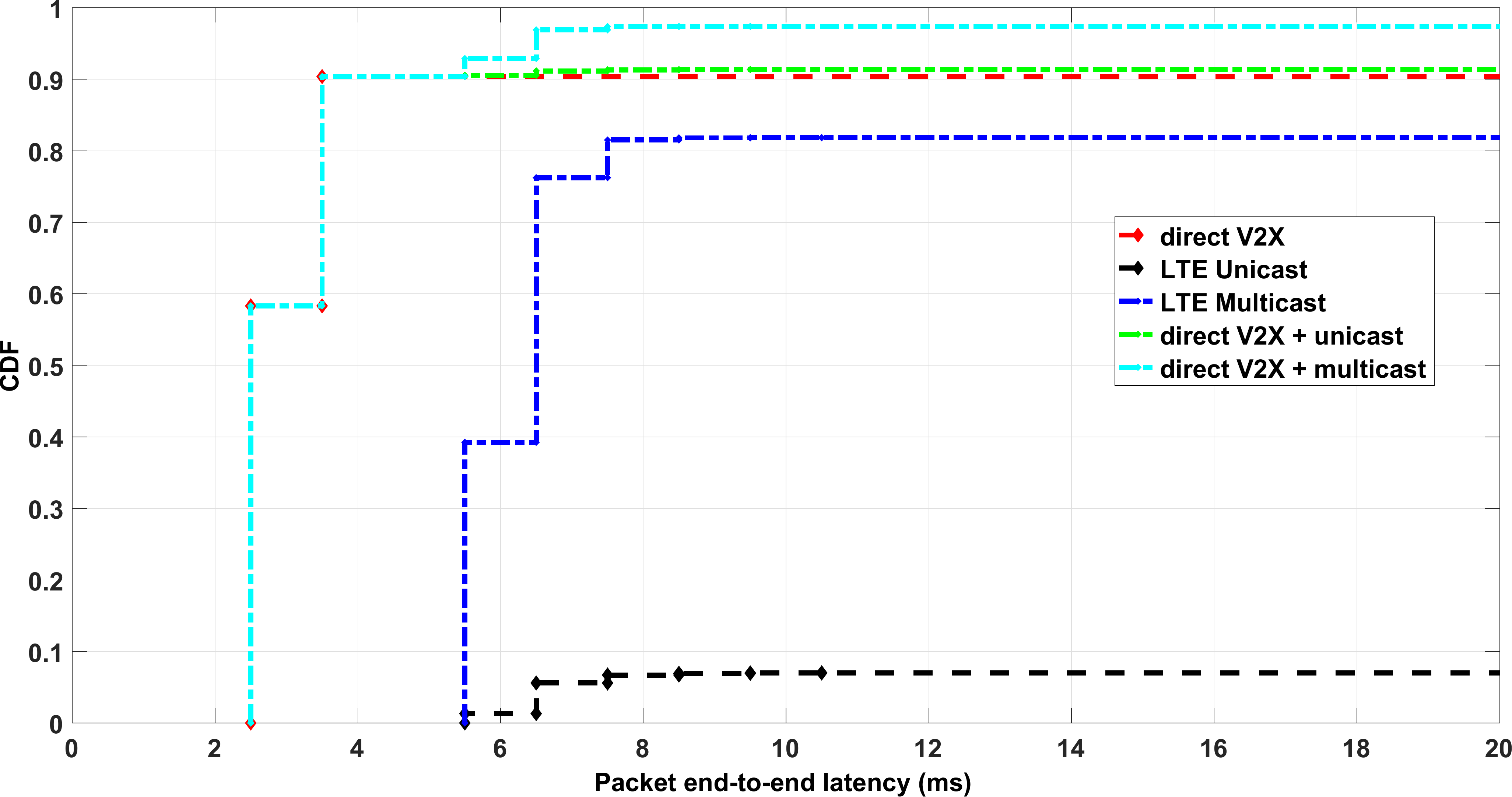}
\caption{CDF of packet E2E latency (Target range = 200 m and 1000 UEs/$\text{km}^2$)}
\label{200m}
\end{figure}
\begin{figure}
\includegraphics[width=3.4in]{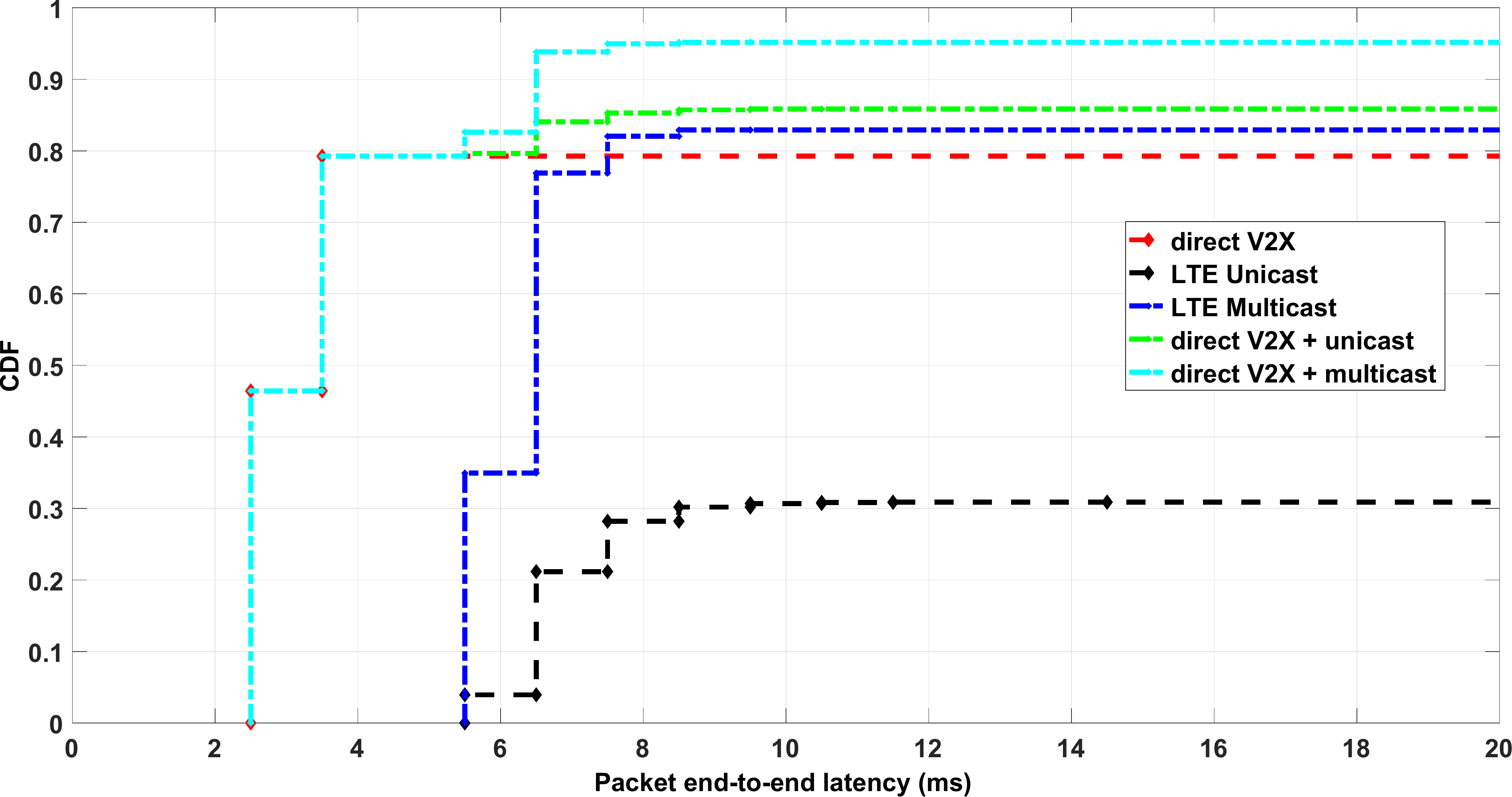}
\caption{CDF of packet E2E latency (Target range = 300 m and 500 UEs/$\text{km}^2$)}
\label{300m}
\end{figure}
\begin{figure}
\includegraphics[width=3.4in]{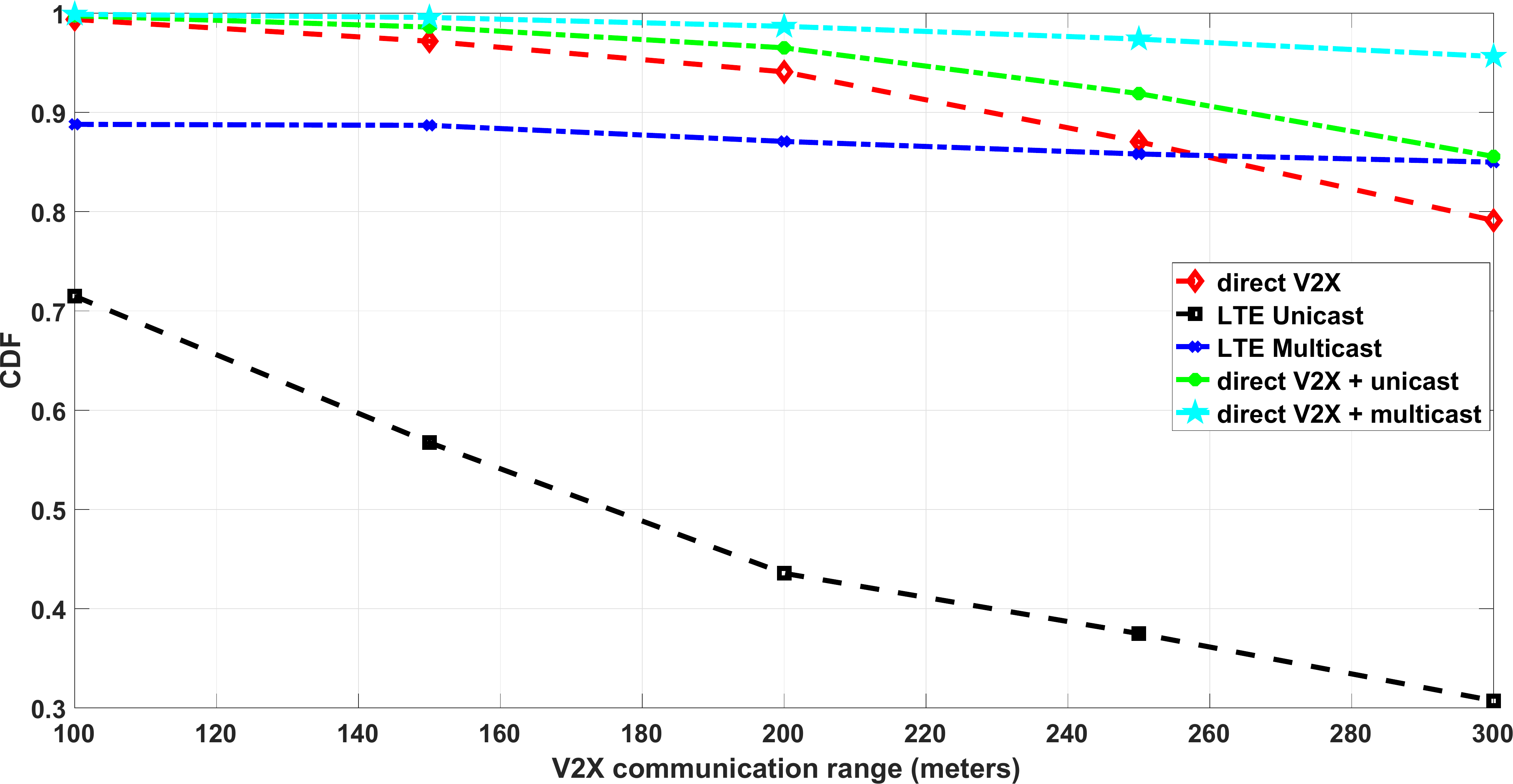}
\caption{PRR of different technologies w.r.t. the different communication ranges (500 UEs/$\text{km}^2$)}
\label{500UE}
\end{figure}

In this part, the simulation results obtained from a system level simulator are provided and the LTE technology is applied to model the radio link performance. The evaluation methodology of packet E2E latency over PC5 and LTE-Uu interfaces are introduced in \cite{V2X} and \cite{C-V2X}, respectively. Moreover, since our focus in this work is on the RAN, the message transition latency among different BSs is not inspected in detail. It is assumed that both the CP and UP functionalities of the BM-SC and MBMS-GW are localized at the edge of the RAN \cite{23785} and a latency value of 1 ms is assumed for a message transition among different BSs. Moreover, as mentioned in Sect.~\ref{LTE-C} and Sect.~\ref{sidelink}, both the eMBMS and the sidelink communication for V2X services correspond to a multicast transmission mode. Thus, a fixed MCS with resource efficiency of 0.6016 bit/Hz (i.e., corresponding to channel quality indicator (CQI) index 4 in LTE network) is applied for the sidelink transmission and another fixed MCS with resource efficiency of 0.887 bit/Hz (i.e., corresponding to CQI index 5 in LTE network) is used for eMBMS.\\
In Fig.~\ref{200m}, the cumulative distribution function (CDF) of the packet E2E latency is plotted where the target V2X communication range is set to be 200 meters and a vehicle density of 1000 vehicles/$\text{km}^2$ is assumed. Please note that the CDF curves do not converge to 100\% due to the fact that certain packets are not successfully received at Rxs. In this figure, the performance of the direct V2X communication over PC5 and the performance of the LTE-Uu interface by using both the unicast and multicast transmission modes in downlink are provided. As can be seen, the LTE unicast mode has the worst performance due to its large resource requirement. In comparison, the LTE multicast in downlink is more resource-efficient in the considered V2X scenario and therefore it has a better performance w.r.t. the packet latency and the PRR. As mentioned before, V2X communication refers to a local information exchange and therefore the direct V2X communication over PC5 can provide a good performance within a moderate communication range. This point is also illustrated in Fig.~\ref{200m}, as the PRR for the direct V2X communication is higher than both the LTE unicast and multicast schemes. In addition to that, since the data packet does not go through the network infrastructure in the direct V2X communication, the packet E2E latency is shorter than the other two schemes \cite{D11}. In order to improve the reliability, the performances of two multi-RATs schemes are also given. In the first multi-RATs scheme, the V2X packets are transmitted over both the PC5 and the LTE-Uu unicast interfaces. And its PRR is better than the case if the packets travel through a single-RAT. However, as the LTE-Uu unicast provides a comparably low PRR, the improvement from the multi-RATs is very slight compared with the performance of the direct V2X communication. In the second multi-RATs scheme, both sidelink and LTE-Uu multicast are exploited. As can be seen from that curve, this multi-RATs scheme provides a large gain w.r.t. the PRR and it is improved from 90\% to 97\%.\\
In Fig.~\ref{300m}, the V2X communication range is increased to 300 meters and the vehicle density is decreased to 500 vehicles/$\text{km}^2$. As can be seen from this figure, the performance of both the LTE-Uu multicast and unicast transmission schemes are better than the ones shown in Fig.~\ref{200m}, due to a lower overall traffic volume. However, the PRR of the direct V2X communication is worse than that in Fig.~\ref{200m}. This is due to a larger communication range and the V2X Rxs located far from the Tx experience bad radio conditions. Thus, the LTE-Uu multicast transmission outperforms the direct V2X communication in this specific case. Additionally, both the LTE multicast and direct V2X communication have a PRR lower than 85\%. Again, by applying the multi-RATs scheme (i.e., direct V2X + LTE-Uu multicast), the PRR can be efficiently improved to be above 95\%.\\
Finally, the PRRs of the different schemes w.r.t. different communication ranges are plotted (i.e., from 100 meters to 300 meters with a step-width of 50 meters) in Fig.~\ref{500UE}. The vehicle density is set to be 500 vehicles/$\text{km}^2$. As can be seen, the LTE unicast scheme is very sensitive to the different communication ranges, since the range determines the number of Rxs for each V2X packet and therefore a larger communication range requires more resource. At the same time, the performance of the direct V2X communication is also influenced by the communication range as a large transmission distance statistically introduces a bad radio condition. In contrast, the LTE-Uu multicast scheme is less sensitive to the communication range, as the signal propagation distance between a V2X UE and its serving BS is independent of the communication range. Comparing the two multi-RATs schemes, the combination of using LTE-Uu multicast and direct V2X shows a better performance than using LTE-Uu unicast with direct V2X. Last but not least, the performance gain w.r.t. the PRR is more outstanding with an increased target communication range.
\section{Conclusion}\label{con}
In our work, we have introduced the different V2X communication technologies (i.e., LTE-Uu and PC5). In particular, the unicast and multicast transmission modes of LTE-Uu interface have been described with a focus on applying them to V2X communication. Besides, as the PC5 interface is standardized in 3GPP to enable the proximity services between two nearby devices, its application in the direct V2X communication has been also introduced. In order to meet the high reliability requirement, we have proposed a multi-RATs scheme where the two different radio access technologies are combined. In addition, a system level simulator has been implemented to show the performance of different technologies. From the simulation results, it can be seen that the performance of the different technologies is related to the concrete scenario and the proposed multi-RATs scheme efficiently improves the service quality.\\
\section*{Acknowledgment}
A part of this work has been supported by the Federal Ministry of Education and Research of the Federal Republic of Germany (BMBF) in the framework of the project 5G-NetMobil with funding number 16KIS0692. The authors would like to acknowledge the contributions of their colleagues, although the authors alone are responsible for the content of the paper which does not necessarily represent the project.


%

\end{document}